# A mildly relativistic wide-angle outflow in the neutron star merger GW170817

K. P. Mooley (1,2,3,19), E. Nakar (4), K. Hotokezaka (5), G. Hallinan (3), A. Corsi (6), D.A. Frail (2), A. Horesh (7), T. Murphy (8,9), E. Lenc (8,9), D.L. Kaplan (10), K. De (3), D. Dobie (8,9,11), P. Chandra (12,13), A. Deller (14), O. Gottlieb (4), M.M. Kasliwal (3), S. R. Kulkarni (3), S.T. Myers (2), S. Nissanke (15), T. Piran (7), C. Lynch (8,9), V. Bhalerao (16), S. Bourke (17), K.W. Bannister (11), L.P. Singer (18) (Affiliaitions: (1) Hintze Fellow. Oxford, (2) NRAO, (3) Caltech, (4) Tel Aviv University, (5) Princeton University, (6) Texas Tech University, (7) The Hebrew University of Jerusalem, (8) University of Sydney, (9) CAASTRO, (10) University of Wisconsin - Milwaukee, (11) ATNF, CSIRO, (12) NCRA, (13) Stockholm University, (14) Swinburne University of Technology, (15) Radboud University, (16) IIT Bombay, (17) Chalmers University of Technology, (18) NASA GSFC, (19) Jansky Fellow, NRAO/Caltech)

**GW170817 is the first gravitational wave detection of a binary neutron star merger[1]. It was accompanied by radiation across the electromagnetic spectrum and localized[2] to the galaxy NGC 4993 at a distance of 40 Mpc. It has been proposed that the observed gamma-ray, X-ray and radio emission is due to an ultra-relativistic jet launched during the merger, directed away from our line of sight[3,4,5,6]. The presence of such a jet is predicted from models positing neutron star mergers as the central engines driving short-hard gamma-ray bursts[7,8] (SGRBs). Here we show that the radio light curve of GW170817 has no direct signature of an off-axis jet afterglow.**

**While we cannot rule out the existence of a jet pointing elsewhere, the observed gamma-rays could not have originated from such a jet. Instead, the radio data requires a mildly relativistic wide-angle outflow moving towards us. This outflow could be the high velocity tail of the neutron-rich material dynamically ejected during the merger or a cocoon of material that breaks out when a jet transfers its energy to the dynamical ejecta. The cocoon scenario can explain the radio light curve of GW170817 as well as the gamma-rays and X-rays (possibly also ultraviolet and optical emission)[9,10,11,12,13,14,15], and hence is the model most consistent with the observational data. Cocoons may be a ubiquitous phenomenon produced in neutron star mergers, giving rise to a heretofore unidentified population of radio, ultraviolet, X-ray and gamma-ray transients in the local universe.**

The radio discovery[12] of GW170817, as well as observations within the first month post-merger, were interpreted in the framework of classical off-axis jet, cocoon, and dynamical ejecta. We continued to observe GW170817 with the Karl G. Jansky Very Large Array (VLA), the Australia Telescope Compact Array (ATCA) and the upgraded Giant Metrewave Radio Telescope (uGMRT), spanning the frequency range 0.6-18 GHz, whilst optical and X-ray telescopes were constrained by proximity to the Sun. Our radio detections span up to 107 days post-merger (Figure 1 and Methods). These data show a steady rise in the radio light curve and a spectrum consistent with optically-thin synchrotron emission. A joint temporal and spectral power-law fit to these data of the form $S \propto \nu^\alpha t^\delta$, is well-described by a spectral index α=-0.6 and a temporal index δ=+0.8 (see Methods). On 2017 November

18 (93 days post-merger) the peak luminosity at 1.6 GHz was 2 x $10^{27}$ erg $s^{-1}$ $Hz^{-1}$, a luminosity undetectable for even the nearest SGRB afterglow discovered to date[16].

The (sub-luminous) gamma-ray emission detected immediately after the gravitational wave detection[17] must have been emitted by a relativistic outflow[14], but an on-axis jet (scenario A in Figure 2) was ruled out by the late turn-on of the X-ray and radio emission[3,4,5,6,11,12,13]. If GW170817 produced a regular (luminous) SGRB pointing away from us, then the interaction of the jet with the circum-merger medium would have decelerated the jet, and the afterglow emission would have eventually entered into our line of sight, thus producing a so-called off-axis afterglow[18,19]. For this geometry, the light curve rises sharply and peaks when the jet Lorentz factor $\gamma \sim 1/(\theta_{obs}-\theta_j)$, and then undergoes a power law decline ($\theta_{obs}$ is the angle between the jet axis and the line of sight, and $\theta_j$ is the jet opening angle). This behavior is clearly inconsistent with the full light curve shown in Figure 1. The rise is less steep than an off-axis jet and it is consistent with a monotonic increase without either a plateau or a subsequent decay. Initial off-axis models (based on available X-ray and radio data at the time) predicted a radio flux density[3,4,5,12] of ~10 µJy (between 3 GHz and 10 GHz) ~100 days post-merger, while our measured values are at least a factor of five larger. The discrepancy with the off-axis jet model is further demonstrated in Figure 3 where various jet and medium parameters are considered, showing in all cases a similar general light curve shape which cannot fit the data. We have considered a wide range in the phase space of off-axis models, and can rule out an off-axis jet (scenario B in Figure 2) as the origin of the radio afterglow of GW170817. We show below that even if we consider a "structured jet", in which the outflow has an angular dependence of the Lorentz factor and energy (scenario E in Figure 2 represents one such configuration), the observed radiation arises predominantly from a mildly relativistic outflow moving towards us (at an angle less than $1/\chi$), and we do not detect the observational signature of a relativistic core within the structured jet.

With a highly collimated off-axis jet ruled out, we next consider spherical or quasi-spherical ejecta components. A single spherical shell of expanding ejecta will produce a light curve that rises as $S \sim t^3$. The light curve of GW170817 immediately rules out such a simple single-velocity ejecta model. The gradual but monotonic rise seen in our radio data ($S \propto t^{0.8}$; Figure 1) points instead to *on-axis* emission from a mildly relativistic blast wave where the energy is increasing with time (due to more mass residing in slower ejecta, which is seen at later times). For example, using canonical microphysical parameters ($\epsilon_B=0.01$, $\epsilon_e=0.1$), a density of $10^{-4}$ cm$^{-3}$ implies that between day 16 to day 107 the blast wave decelerates from $\chi \sim 3.5$ to $\chi \sim 2.5$ and its isotropic equivalent energy increases from $\sim 10^{49}$ erg to $\sim 10^{50}$ erg. On the other hand, a density of 0.01 cm$^{-3}$ implies a velocity range of 0.8c to 0.65c and energy that rises from $10^{48}$ erg to $10^{49}$ erg. Figure 4 shows that a quasi-spherical outflow with a velocity profile $E(>\beta\chi) \propto (\beta\chi)^{-5}$ provides an excellent fit to the data (see Methods), and it is almost independent of the assumed circum-merger density and microphysical parameters. The energy injection into the blast wave during the time span of the observations (day 16 to day 107) increases its energy by a factor of ~10. The possible origin of the outflow depends on its energy and velocity. A faster and more energetic outflow, with $\chi \sim 2-3$ and energy of $10^{49}$-$10^{50}$ erg, is a natural outcome of the cocoon driven by a wide-angle choked jet[9,11,14] (scenario C in Figure 2). This scenario explains many of the puzzling characteristics of GW170817. First, the breakout of the cocoon from the ejecta can produce the observed sub-luminous gamma-ray signal, including its peak energy and spectral evolution[14] (see also Methods). Second, it provides a natural explanation for the high velocities of the bulk of the ejecta (~0.3c) and for the early bright UV and optical light[11,13,15]. On the other hand, a slower and less energetic outflow, with $\beta \approx 0.8$-$0.6$ ($\gamma$~1.67-1.25) and energy of $10^{48}$-$10^{49}$ erg can arise from the fast tail of the merger ejecta[20,21,22] (scenario D in Figure 2), although we note that this component cannot explain the gamma-ray signal (GRB 170817A) from GW170817. These two scenarios can be easily distinguished by Very Long Baseline Interferometry or monitoring of the radio evolution on ~years timescale.

A *hidden* jet, which does not contribute significantly to the observed afterglow, may still exist (scenario E in Figure 2), but its properties are tightly constrained. First, its edge must be far enough from the line-of-sight ($\gtrsim$10 degrees), which rules out off-axis gamma-ray emission as the source of GRB 170817A. Second, for every reasonable set of parameters, an off-axis jet would have been brighter than the fast tail of the ejecta, implying that the observed emission must be dominated by a $\gamma$~2-3 outflow (i.e. a cocoon) for the jet to remain undetected. In addition, the jet energy should, most likely, be much lower than that of the cocoon, which needs fine-tuning of the jet properties (see Methods). We therefore conclude from the lack of a signature from an off-axis jet, that the jet was likely choked (scenario C in Figure 2).

We compared the 3 GHz radio and X-ray[4,5,6] detections obtained on 2017 September 02-03 (15-16 days post-merger). The measurements at these two disparate frequencies imply a spectral index of -0.6, consistent with our multi-epoch, multi-frequency, radio-only measurements (see Methods and Extended Data Figure 4). It is therefore likely that the radio and X-rays originate from the same (synchrotron) source, viz. a mildly relativistic outflow. This common origin can be confirmed if the X-ray flux continues to rise in a similar manner as the radio. We also highlight that, while at early times the cooling break will lie well above the soft X-ray frequencies, beyond ~$10^2$-$10^3$ days post-merger this break may be seen moving downwards in frequency within the electromagnetic spectrum. If the cooling break stays above $10^{18}$ Hz, the common origin of the radio and X-rays implies that the Chandra telescope will detect a brighter X-ray source (flux between $0.7 \times 10^{-14}$ and $5.2 \times 10^{-14}$ erg cm$^{-2}$ s$^{-1}$ in the 0.3-10 keV band; see Methods) during its observation of GW170817 on December 03-06 (*note: subsequent to the submission of this paper, the X-ray observations took place and confirmed our prediction*). If a different spectral index is derived from these X-ray observations relative to the in-band radio spectral index presented here, or indeed at any time within ~1000 days of the merger, it will indicate that the cooling break has already shifted below the X-ray band, which would favor the fast tail of the merger ejecta as the common source of the X-ray and radio emission (see Methods).

The confirmation of a wide-angle outflow in GW170817 bodes well for electromagnetic counterpart searches of future gravitational wave events. Although on-axis (and slightly off-axis; $\theta_{obs}$<20 degrees) jets produce bright panchromatic afterglows, they represent only a small fraction (~10%) of the gravitational wave events (factoring in the larger detectable distance for face-on events[23]). In contrast, the emission from wide-angle cocoons[9,10,11] will be potentially seen in a much larger fraction of events, and at virtually all wavelengths, thus increasing the probability of the detection of electromagnetic counterparts. The radio emission from the cocoon, evolving on timescales of weeks to months, especially provides a distinct signature (as opposed to the more common supernovae and AGN transients) and diagnostics for observers. Specifically in the case of GW170817, continued monitoring of the radio light curve will provide an independent constraint on the circum-merger density and thereby the properties of the blast wave that dominated the early-time radio emission.

Our radio data support the hypothesis of a choked jet giving rise to a mildly relativistic cocoon (scenario C in Figure 2), but this is only one of the possible outcomes of neutron star merger events (see Figure 2). In some cases, the jet may break out after depositing a fraction of its energy into the cocoon, thereby still successfully producing a SGRB[11] (scenario E in Figure 2). Indeed, a plateau in the distribution of SGRB durations has been highlighted as evidence that SGRB jets often propagate through slower traveling ejecta before breakout and at times it is choked[24]. The relative fractions of neutron star mergers that successfully produce a SGRB or a choked jet can be directly probed via radio follow-up of a sample of neutron star mergers in the upcoming LIGO-Virgo campaigns.

**Acknowledgements.** We would like to acknowledge the support and dedication of the staff of the National Radio Astronomy Observatory and particularly thank the VLA Director, Mark McKinnon, as well as Amy Mioduszewski and Heidi Medlin, for making the VLA campaign possible. We thank Britt Griswold (NASA/GSFC) for beautiful graphic arts. SK thanks Mike Shull for discussions. We thank the anonymous referees for their comments. The National Radio Astronomy Observatory is a facility of the National Science Foundation operated under cooperative agreement by Associated Universities, Inc. We thank the GMRT staff for scheduling our observations. The GMRT is run by the National Centre for Radio Astrophysics of the Tata Institute of Fundamental Research. The Australia Telescope Compact Array is part of the Australia Telescope National Facility which is funded by the Australian Government for operation as a National Facility managed by CSIRO. KM's research is supported by the Hintze Centre for Astrophysical Surveys which is funded through the Hintze Family Charitable Foundation. EN acknowledges the support of an ERC starting grant (GRB/SN) and an ISF grant (1277/13). GH acknowledges the support of NSF award AST-1654815. AC acknowledges support from the National Science Foundation CAREER award #1455090 titled "CAREER: Radio and gravitational-wave emission from the largest explosions since the Big Bang". AH acknowledges support by the I-Core Program of the Planning and Budgeting Committee and the Israel Science Foundation. TM acknowledges the support of the Australian Research Council through grant FT150100099. Parts of this research were conducted by the Australian Research Council Centre of Excellence for All-sky Astrophysics (CAASTRO), through project number CE110001020. DK was supported by NSF grant AST-1412421. MK's work was supported by the GROWTH (Global Relay of Observatories Watching Transients Happen) project funded by the National Science Foundation under PIRE Grant No 1545949. This work is part of the research program Innovational Research Incentives Scheme (Vernieuwingsimpuls), which is financed by the Netherlands Organization for Scientific Research through the NWO VIDI Grant No. 639.042.612-Nissanke and NWO TOP Grant No. 62002444--Nissanke. PC acknowledges support from the Department of Science and Technology via SwarnaJayanti Fellowship awards (DST/SJF/PSA-01/2014-15). TP acknowledges the support of Advanced ERC grant TReX. VB acknowledges the support of the Science and Engineering Research Board, Department of Science and Technology, India, for the GROWTH-India project.

**Author Contributions.** KM, EN, KH, GH and DF wrote the paper. AC compiled the references. AC and AH compiled the methods section. DD and KD compiled the radio measurements table. KM managed the VLA observing program and processed all the VLA data. SM, AD and SB helped plan the VLA observations. EN, KH, DK and KM prepared the figures. TM planned and managed ATCA observations and data analysis and contributed to the manuscript text. DK helped propose for and plan the ATCA observations and contributed to the manuscript text. EL, DD, CL and KB helped with ATCA observations and data reduction. KD planned and managed GMRT observations and contributed to manuscript text. KM and PC processed the GMRT data. VB helped in the GMRT observations. OG and EN provided the cocoon simulation. KH provided the spherical ejecta model. SN did the GW and cocoon rates analysis. SK, TP, MK and LS provided text for the paper. All coauthors discussed the results and provided comments on the manuscript.

**Competing Interests.** The authors declare that they have no competing financial interests.

**Correspondence.** Correspondence and requests for materials should be addressed to K.P.M. (email: kunal@astro.caltech.edu).

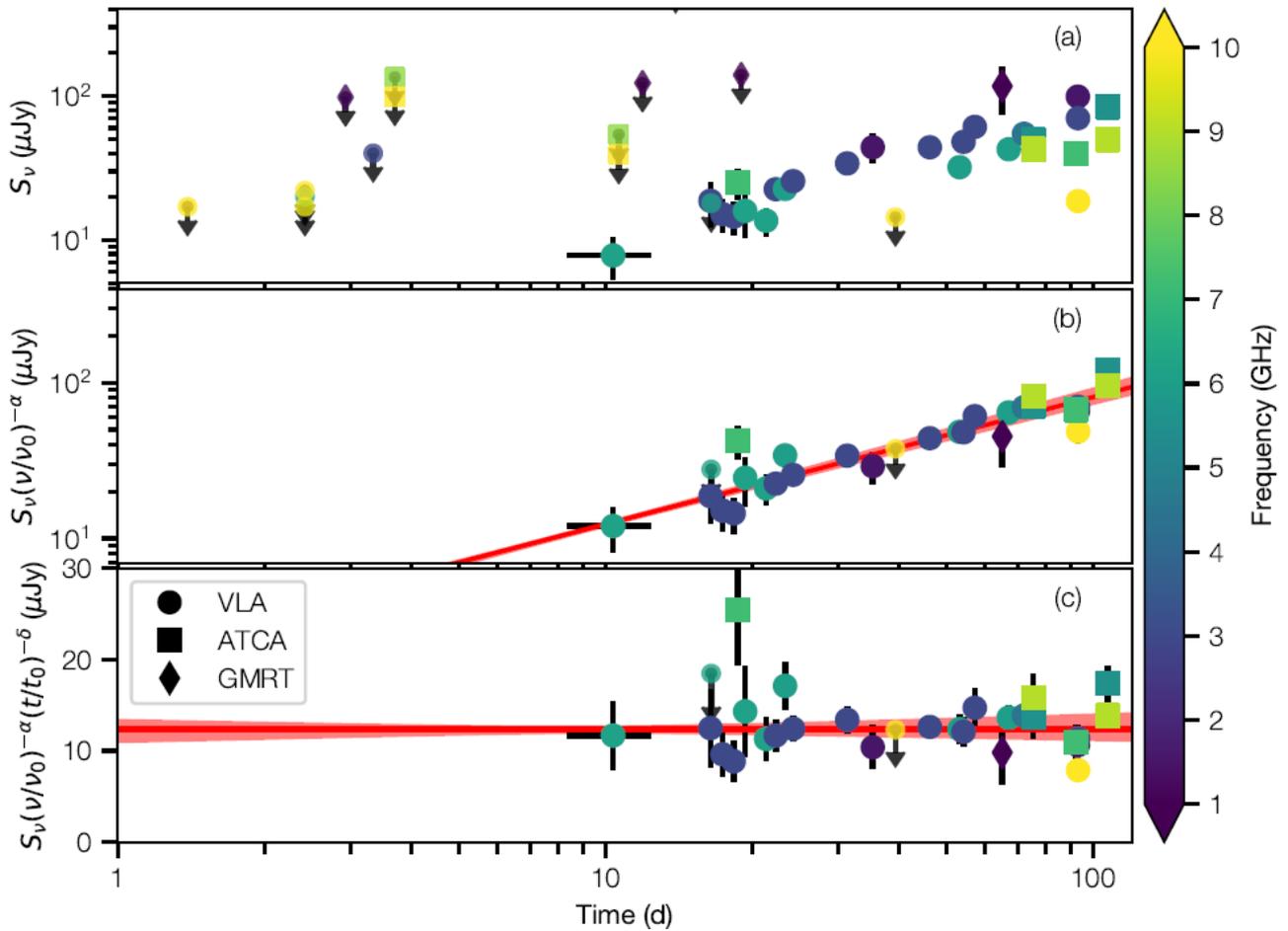

**Figure 1. The radio light curve of GW170817.**

Panel (a): The flux densities corresponding to the detections (markers with 1σ error bars; some data points have errors smaller than the size of the marker) and upper limits (markers with downward-pointing arrows) of GW170817 at frequencies ranging from 0.6-15 GHz between day 16 and day 107 post-merger (ref. 12 and Extended Data Table 1). Panel (b): Same as the panel (a) but with flux densities corrected for the spectral index α=-0.61 (see Methods) and early-time, non-constraining, upper limits removed. The fit to the light curve with the temporal index δ=0.78 (see Methods) is shown as a red line and the uncertainty in δ (+/-0.05) as the red shaded region. Panel (c): Residual plot after correcting for the spectral and temporal variations. The observing frequencies are color coded according to the colorbar displayed at the right (black for ≤1 GHz and yellow for ≥10 GHz). The marker shapes denote measurements from different telescopes.

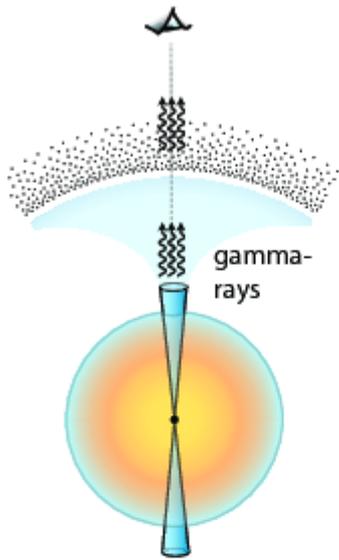

A. On-axis Jet
SGRB and afterglow
(RULED OUT)

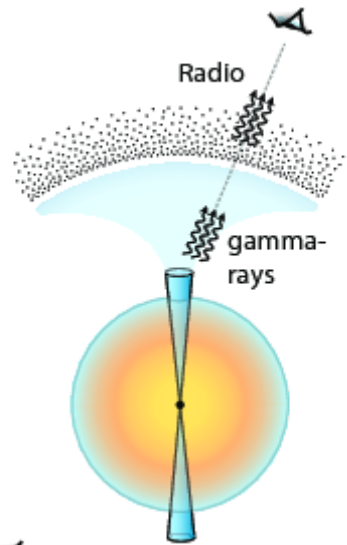

B. Off-axis Jet
SGRB and afterglow
(RULED OUT)

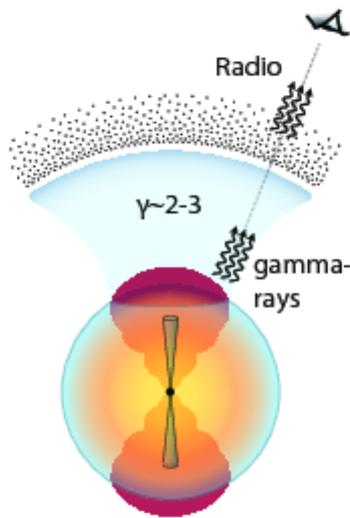

C. Choked Jet
Cocoon gamma-rays
and afterglow
(Most likely)

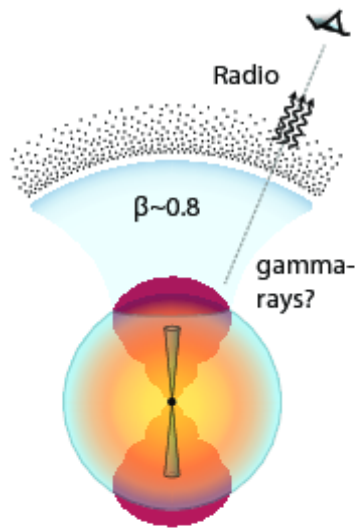

D. Choked Jet
Fast ejecta afterglow
(Less likely)

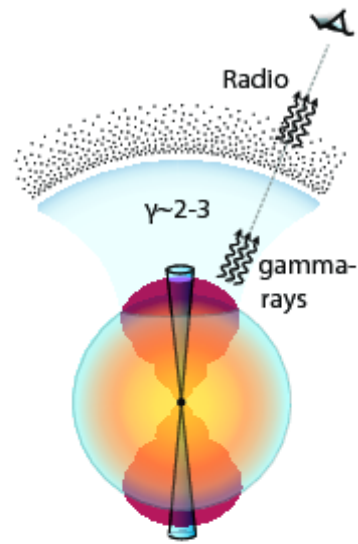

E. Successful hidden Jet
Cocoon gamma-rays
and afterglow
(Less likely)

**Figure 2. Schematic illustration of the various possible jet and dynamical ejecta scenarios in GW170817.**

A) A jet seen on-axis, generating both the low-luminosity gamma-rays and the observed radio afterglow. This scenario cannot explain the late rise of the radio emission. It is also unable to explain[11] how a low-luminosity jet penetrates the ejecta. It is therefore ruled out. B) A regular (luminous) SGRB jet seen off-axis, producing the gamma-rays and the radio. The continuous moderate rise in the radio light curve rules out this scenario. C) A choked jet giving rise to a mildly relativistic ($\gamma$~2-3) cocoon which generates the gamma-rays and the radio waves via on-axis emission. This is the model that is most consistent with the observational data. It accounts for the observed gamma-rays, X-rays (possibly also the ultraviolet and optical emission) and the radio emission, and provides a natural explanation for the lack of an off-axis jet signature in the radio. D) The fast velocity tail ($\beta$~0.8-0.6c, i.e. $\gamma$~1.67-1.25) of the ejecta produces the radio emission. In this case, the jet must be choked (otherwise its off-axis emission should have been seen). While the radio emission is fully consistent with this scenario, the energy deposited in faster ejecta ($\gamma$~2-3) must be very low. In this scenario, the source of the observed gamma-rays remains unclear. E) A successful jet that drives a cocoon but does not have a clear signature in the radio. The cocoon generates the gamma-rays and the radio emission, and outshines the jet at all wavelengths. This scenario is less likely based on theoretical considerations, which suggest that the jet and the cocoon should have comparable energies, in which case the jet signature would have been observed in the radio band. This scenario can also be visualized as a "structured" jet, having a relativistic narrow core surrounded by a mildly relativistic wide-angle outflow, in which an off-axis observer does not see any signature of the core. The relativistic core could have produced a regular SGRB for an observer located along the axis of the jet. Such a jet, if it exists, could be too weak (made a sub-dominant contribution to the radio light curve early on) or too strong (such that its radio and X-ray signatures will be observed in the future; see Methods).

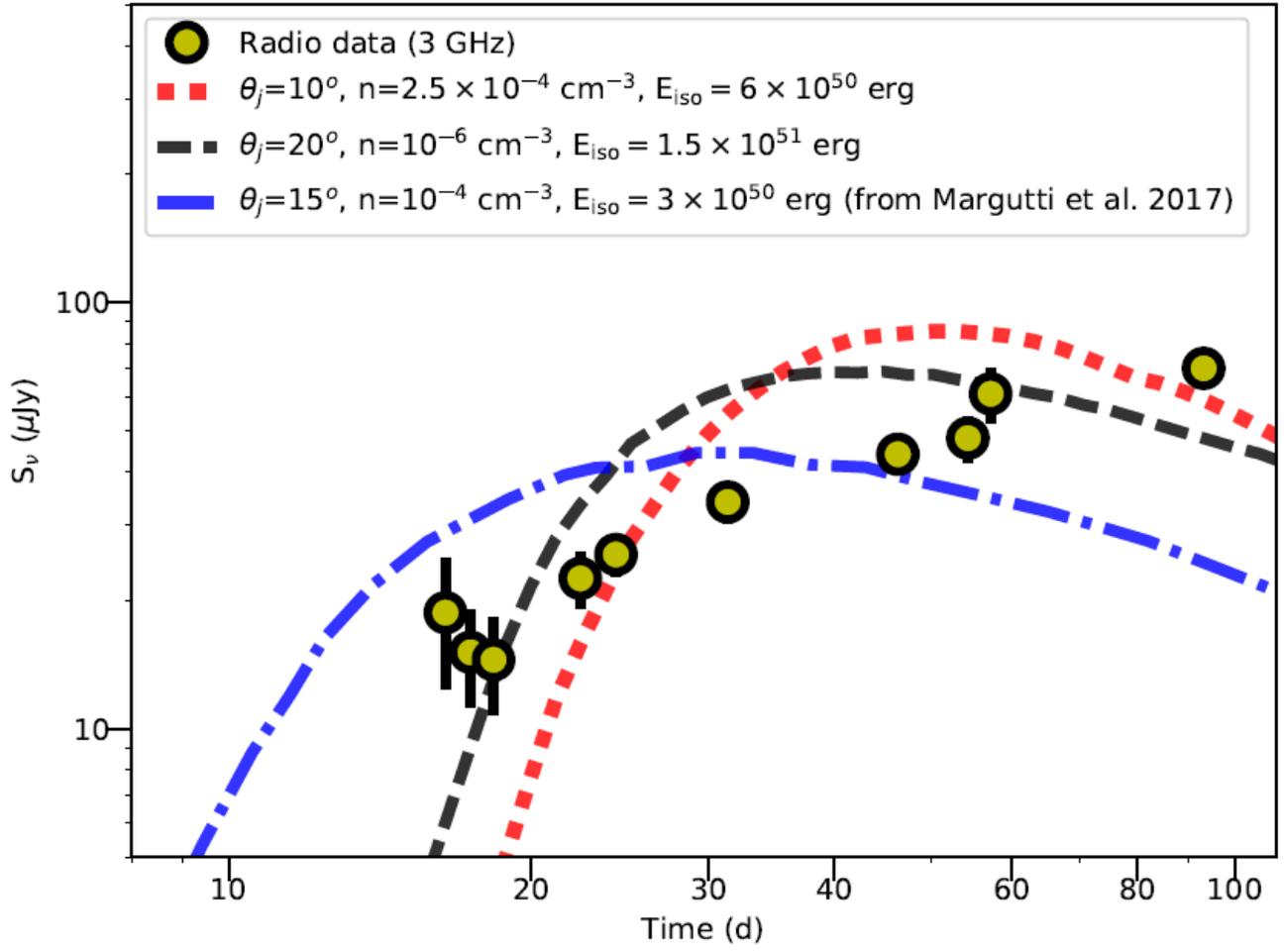

**Figure 3. Off-axis jet models.**

Synthetic light curves with a range of jet opening angles $\theta_j$, isotropic-equivalent energy $E_{iso}$, and the ISM density n (see Methods) overplotted on the 3 GHz light curve (error bars are 1σ; ref. 12 and Extended Data Table 1). The overall shape of the light curve remains unchanged even after changing these parameters. We have considered a wide range of parameters in the phase space of off-axis models (including unlikely scenarios like $n=10^{-6}$ cm$^{-3}$; see Methods); none of the models give a good fit to the observational data, and hence we rule out the classical off-axis jet scenario as a viable explanation for the radio afterglow. The dashed black and dotted red curves are calculated using the codes described in the Methods. The dashed-dot blue curve is taken from figure 3 of ref. 4

(scaled to 3 GHz using α=-0.6). All off-axis models assume $\theta_{obs}$ = 26 deg, $\epsilon_e$ = 0.1, $\epsilon_B$ = 0.01 and p=2.2. (see main text and Methods).

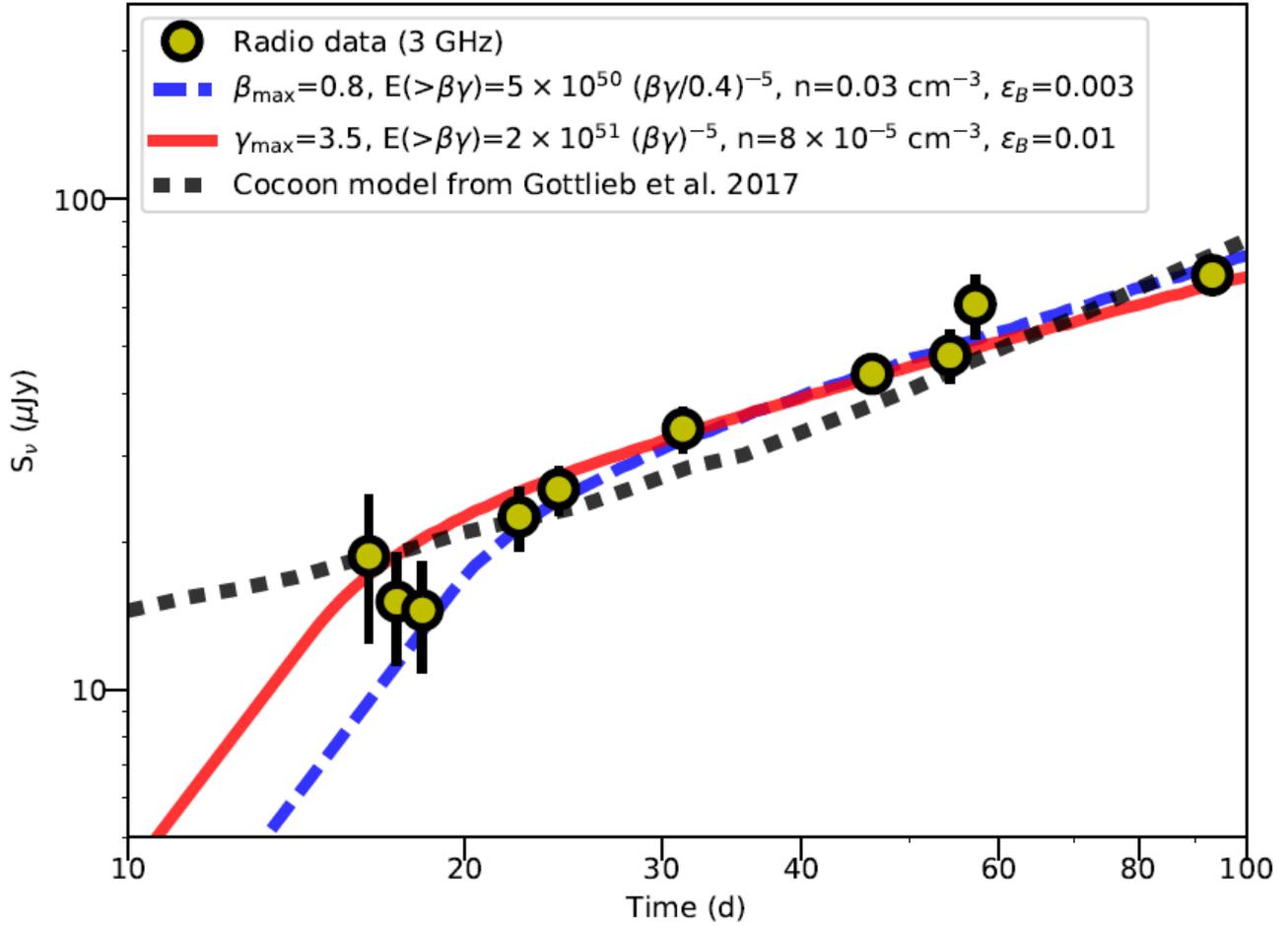

**Figure 4. Quasi-spherical ejecta models.**

Radio light curves arising from quasi-spherical ejecta with velocity gradients, overplotted on the 3 GHz data spanning days 16-93 post-merger (filled yellow circles; error bars are 1σ; ref. 12 and Extended Data Table 1). The solid red and dashed blue light curves represent power law models with maximum Lorentz factors $\gamma=3.5$ and $\gamma=1.67$ respectively (i.e. maximum $\beta=v/c=0.96$ and 0.8 respectively). These curves approximately correspond to the cocoon and dynamical ejecta, respectively. The shallow rise of the radio light curve is consistent with a profile $E(>\beta\gamma) \propto (\beta\gamma)^{-5}$. For $n\sim0.03$ cm$^{-3}$, the observed radio emission at 93 days is produced by an ejecta component with a velocity of $\sim0.6c$ and kinetic energy of $\sim10^{49}$ erg. For a lower ISM density, $\sim10^{-4}$ cm$^{-3}$, the radio emission at 93 days is produced by a component with a velocity of $0.9c$ and energy $10^{50}$ erg. Parameters $\epsilon_e=0.1$ and $p=2.2$ are used for both models. Also shown for reference is the cocoon model light curve (dotted black curve) taken from ref. 14, where parameter values n=1.3x10$^{-4}$ cm$^{-3}$, $\epsilon_B$=0.01, $\epsilon_e$=0.1 and p=2.1 are used.

**Methods**

**1. Radio Data Analysis.**

**VLA**. Radio observations of the GW170817 field were carried out with the Karl G. Jansky Very Large Array in its B configuration, under a Director Discretionary Time (DDT) program (VLA/17B-397; PI: K. Mooley). All observations were carried out with the Wideband Interferometric Digital Architecture (WIDAR) correlator in multiple bands including L-band (nominal center frequency of 1.5 GHz, with a bandwidth of 1 GHz), S-band (nominal center frequency of 3 GHz, with a bandwidth of 2 GHz), and C-band (nominal center frequency of 6 GHz, with a bandwidth of 4 GHz). We used QSO J1248-1959 (L-band and S-band) and QSO J1258-2219 (C-band) as our phase calibrator sources, and 3C 286 or 3C 147 as flux density and bandpass calibrators. The data were calibrated and flagged for radio frequency interference (RFI) using the VLA automated calibration pipeline which runs in the Common Astronomy Software Applications package (CASA[25]). We manually removed further RFI, wherever necessary, after calibration. Images of the observed field were formed using the CLEAN algorithm (with the "psfmode" parameter set to Hogbom[26]), which we ran in the interactive mode. The results of our VLA follow-up campaign of GW170817 are reported in Extended Data Table 1, and the image cutouts are shown in Extended Data Figure 1. The flux densities were measured at the Gaia/HST position[27]. Flux density measurement uncertainties denote the local root-mean-square (rms) noise. An additional 5% fractional error on the measured flux density is expected due to inaccuracies in the flux density calibration. For non-detections, upper-limits are calculated as three times the local rms noise in the image.

**ATCA**. We observed GW170817 on 2017 November 01, November 18 and December 02 using the Australia Telescope Compact Array (ATCA) under a target of opportunity program (CX391; PI: T. Murphy). During these observations the array was in configurations 6A, 1.5C and 6C respectively. We observed using two 2 GHz frequency bands with central frequencies of 5.5 and 9.0 GHz. For both epochs, the flux scale and bandpass response were determined using the ATCA primary calibrator PKS B1934-638, and observations of QSO B1245-197 were used to calibrate the complex gains. The visibility data were reduced using the standard routines in the MIRIAD environment[28]. The calibrated visibility data were split into the separate bands (5.5 GHz and 9.0 GHz), averaged to 32 MHz channels, and imported into DIFMAP[29]. Bright field sources were modeled separately for each band using the visibility data and a combination of point-source and Gaussian components with power-law spectra. With the field sources modelled and subtracted from the visibility data, the dominant emission in the residual image was from GW170817. Restored images for each band were generated by convolving the model components with the restoring beam, adding the residual map and then averaged to form a wide-band image. Image-based Gaussian fitting for an unresolved source was performed in the region of GW170817, leaving the flux density and source position unconstrained. The source position from the fitting agrees with the Gaia/HST position[27] of GW170817. The measured radio flux densities in the combined images are reported in the Extended Data Table 1, and the image cutouts are shown in Extended Data Figure 1.

**GMRT.** We carried out observations of the GW170817 field with the upgraded Giant Meterewave Radio Telescope (uGMRT) at 700 MHz under a DDT program (DDTB288; PI: K. De). All observations were carried out with 400 MHz bandwidth centered at 750 MHz using the non-polar continuum interferometric mode of the GMRT Wideband Backend (GWB[30]). Pointings were centered at the location of the optical transient. 3C 286 was used as the absolute flux scale and bandpass calibrators, while phase calibration was done with the sources J1248-199 (for the 2017 September 16 observation) and 3C 283 (for all other observations). These data were calibrated and RFI flagged using a custom-developed CASA pipeline. The data were then imaged interactively with the CASA task CLEAN, incorporating a few iterations of phase-only self-calibration by building a model for bright sources in the field with each iteration. The GMRT flux density measurements at the Gaia/HST position[27] are reported in the Extended Data Table 1. The image cutouts are shown in Extended Data Figure 1.

**1.1 Radio Data Power-law Fit**

We carried out a least-squares fit to the assembled radio data as a function of time and frequency, using a two-dimensional power-law model:

$S(\nu,t) = S_0 \, (\nu/\nu_0)^\alpha \, (t/t_0)^\delta$

The fit results are shown in Extended Data Figure 2, where we find good results for $\alpha=-0.61+/-0.05$, $\delta=0.78+/-0.05$, $S_0=13.1+/-0.4$ µJy, $\nu_0=3$ GHz and $t_0=10$ d. The fit has $\chi^2=42.3$ for 44 degrees-of-freedom, although there are only 27 detections among the 47 data-points.

**1.2 Multi-epoch radio spectra**

In Extended Data Figure 3 we show the radio continuum spectra obtained at different epochs. All epochs are individually consistent with the spectral index α=-0.61 within 1σ.

**2. Model Descriptions.**

**2.1 Off-axis afterglows.**

The radio light curves were calculated using two independent semi-analytic codes[31,32], which are based on similar approximations. Both codes were compared to, and have been found to be largely consistent with, the light curves produced by the BOXFIT code[33]. In short, both codes approximate the jetted blast wave at any time in the source-frame as a single zone emitting region which is a part of a sphere with an opening angle, $\theta_j$. The hydrodynamics includes the shock location and velocity, and the jet spreading. The hydrodynamic variables in the emitting region are set to their values immediately behind the shock. The emission from each location along the shock is calculated using standard afterglow theory[34], where the microphysics is parameterized by the fraction of internal energy that goes to the electrons, $\epsilon_e$, the fraction of internal energy that goes to the magnetic field, $\epsilon_B$, and the power-law index of the electron distribution. The code calculates the rest frame emissivity at any time and any location along the shock and the specific flux observed at a given viewing angle at a given time and frequency is then found by integrating the contribution over equal-arrival-time surfaces, with a proper boost to the observer frame.

**2.2 Quasi-spherical ejecta.**

Radio light curves arising from quasi-spherical outflows, e.g., a cocoon and the tail of the dynamical ejecta, are approximately described by a model with a single one-dimensional velocity profile: $E(>\beta\gamma) \propto (\beta\gamma)^{-k}$, where $\beta$ is a velocity in units of the speed of light and $\gamma$ is a Lorentz factor. The slope of the observed radio light curve is consistently explained with k=5. The light curves are calculated using the same codes as in section 2.1. In Figure 4, we show two cases: (1) a cocoon model, $E(>\beta\gamma) = 2 \times 10^{51} (\beta\gamma)^{-5}$ erg with a maximum Lorentz factor of 3.5, n=8x10$^{-5}$ cm$^{-3}$, and $\epsilon_B$=0.01, and (2) a dynamical ejecta model, $E(>\beta\gamma) = 5 \times 10^{50} (\beta\gamma/0.4)^{-5}$ erg with a maximum velocity of 0.8c, n=0.03 cm$^{-3}$, and $\epsilon_B$=0.003. This velocity profile of the dynamical ejecta contains a larger mass traveling faster than 0.6c by a factor of ~5 compared with that found in general relativistic numerical simulations[20,21]. The small amount of mass ejected at these high velocities is plausible since the simulations are affected by finite resolution and artificial atmosphere. In addition, Figure 4 shows a prediction from the full 2D simulation of a choked jet and the resulting cocoon presented in ref. 9. The light curve is taken from figure 4 of ref. 14 without any attempt to fit the radio data that was added since it was published. A more detailed publication reporting the full set of 2D simulations is in preparation. FInally, an upper limit on the ISM density[12] of 0.04 cm$^{-3}$ suggests that the ejecta contains a fast moving component with v≳0.6c. For all the models shown in Figure 4, the mass of the ejecta that produces the radio signal up to 93 days is only ~10$^{-5}$ M⊙. This velocity is faster, and the mass is much lower, than those inferred from the kilonova emission[35]. We note that kilonova ejecta will produce observable radio signals on a timescale of years.

## 3. Hiding an off-axis jet

Hiding a luminous off-axis jet (of the type seen in regular SGRBs), given the radio data, is not trivial. First, the jet emission peaks once its Lorentz factor drops to $\sim 1/(\theta_{obs}-\theta_j)$, where $\theta_{obs}$ is the viewing angle with respect to the jet axis and $\theta_j$ is the jet opening angle. Thus, emission from a jet that points only slightly away from us (<10 degrees), will peak when its Lorentz factor is high ($\gtrsim 6$). Since the flux in the radio at a given time is extremely sensitive to the blast wave Lorentz factor (roughly as $\gamma^{10}$) a jet at that angle will be much brighter than any on-axis mildly relativistic outflow around the peak, even if the outflow carried much more energy than the jet. Therefore, a hidden jet must be far away from the line-of sight, namely $\theta_{obs}-\theta_j \gtrsim 10$ degrees. At such angle, any gamma-ray signal produced by a relativistic jet will be too faint to explain the observed gamma-ray signal[11]. Thus, while our previous radio observations strongly disfavored a regular SGRB seen off-axis as the origin of the gamma-rays[12] (scenario B in Figure 2), the additional observations presented here practically rule this out.

The extreme dependence of the radio flux density on the blast wave Lorentz factor also implies that, for reasonable parameters also at $\theta_{obs}-\theta_j \gtrsim 10$ degrees, off-axis jet emission will outshine a blast wave driven by material with $\beta \sim 0.8$ ($\gamma \sim 1.67$). Thus, the radio emission from an off-axis jet may remain undetected only if the observed emission is dominated by an on-axis material with $\gamma \sim 3$, which is most likely a cocoon. In that case, a jet that is far from the line of sight may be hidden in two ways, either by being significantly less energetic than the on-axis outflow or, surprisingly, by being significantly more energetic (scenario E in Figure 2). In the latter case the jet emission will not appear in the radio data available so far if it is so energetic that its Lorentz factor at day 93 is still significantly larger than $\theta_{obs}-\theta_j$. For example, a 10 degree jet with an isotropic equivalent energy of $10^{52}$ erg, that propagates in circum-merger density of $10^{-4}$ cm$^{-3}$ and observed at an angle of 30 degrees, peaks after 200 days and its brightness is comparable to the observed data only around day 90 ($\epsilon_B=0.01$, $\epsilon_e=0.1$). While we cannot rule out this option, the extreme jet energies make it unlikely, but if this is the case then we will see the jet contribution in the future.

The other possibility, that the jet is less energetic than the on-axis outflow (again scenario E in Figure 2), cannot be tested observationally. However, it is unlikely based on theoretical considerations. The energy of the cocoon is distributed over a large range of velocities. Thus, the energy of the mildly relativistic ejecta ($\gamma\sim 3$) is expected to be only a small fraction of the total cocoon energy[9]. Moreover, observationally we see that the energy carried by slower moving on-axis material is at least a factor of 10 larger than energy carried by high velocity on-axis material. Now, the ratio between the total energy in the cocoon and the energy in the jet depends on the ratio between the time spent by the jet in the ejecta before it breaks out and the time over which the jet launching continues after the breakout takes place. The engine that launches the jet is not affected by the propagation of the jet though the ejecta and is causally disconnected from the jet head, if and when it breaks out of the ejecta. Therefore, there is no reason for the engine to stop upon breakout and without fine tuning. If the jet breaks out successfully the launching of the jet is expected to continue over a time that is comparable to or larger than the time it takes for the jet to cross the ejecta. As a result, the energy in the jet is expected to be comparable or larger than that in the cocoon. Thus, it is highly unlikely that the jet is less energetic than the fastest cocoon material, which as noted above carries only a small fraction of the total cocoon energy.

We therefore conclude that there are no probable scenarios in which a jet successfully breaks out, producing an SGRB seen by another (non-Earth) observer, and remains undetected by our radio observations. We find the case in which the jet is choked as the one that provides the best explanation to entire set of observations available to date.

**4. The origin of the gamma-rays**

Since a hidden jet cannot produce the observed gamma-rays and the rising radio emission indicates a mildly relativistic wide-angle outflow moving towards us, we can expect that this outflow is also the origin of the gamma-rays. We do not see any plausible scenario in which the kilonova ejecta can produce the gamma-rays by itself. Compactness arguments imply that this ejecta is too slow[11,14] and there is no natural dissipation process that can convert the kinetic energy of the ejecta to gamma-rays. The cocoon, on the other hand, can produce the gamma-rays. It has sufficient energy and its Lorentz factor is sufficiently high to avoid compactness issues, so in the presence of a dissipation mechanism it can produce the observed gamma-rays[9,10,,36,37]. For example, a breakout of the shock driven by the cocoon through the expanding ejecta can produce the observed signal, accounting for its luminosity, duration, peak energy and spectral evolution[9].

**5. Lower limit to the circum-merger density**

The mean cosmological baryon density is a function of the D/H ratio[38], primordial Helium density[39], cosmographic parameters[40] and the fraction of diffuse baryons in the IGM ($f_{IGM}$) and is given[41] as,

$n_H \sim (1.88 \times 10^{-7} \text{cm}^{-3}) f_{IGM} (1+z)^3$.

We adopt[41] $f_{IGM}=0.7$. At $z\sim0$, a density of $10^{-6}$ cm$^{-3}$ corresponds to a baryon over density $\Delta_b=5$. For the Lyman-alpha forest, $\Delta_b$ is in the range of 10-50, whereas that in condensed halos[41] is $10^2<\Delta_b<10^4$. Thus, in the case of GW170817, a lower limit to the ambient density is $2\times10^{-5}$ cm$^{-3}$ and a typical value[42] would be $\sim10^{-4}$ cm$^{-3}$.

**6. Radio-X-ray comparison**

The 3 GHz flux density measured[12] on 2017 September 03.9 is 15+/-4 µJy. Scaling the X-ray fluxes given in ref. 6 (reported in the energy range 0.3-8 keV) to the values reported in ref. 5 (0.3-10 keV), we estimate the X-ray flux on 2017 September 02.2 as $5.5 \times 10^{-15}$ erg cm$^{-2}$ s$^{-1}$, with a 1σ uncertainty of $\sim1.5 \times 10^{-15}$ erg cm$^{-2}$ s$^{-1}$. We use this information (X-ray flux density is 0.23+/-0.06 nJy at a nominal center frequency of $4\times10^{17}$ Hz) to calculate the spectral index between the radio and X-ray frequencies as -0.60+/-0.03. This is consistent with our estimated value of the radio-only spectral index, -0.61+/-0.05, within 1σ. Therefore the radio emission and X-rays likely originate from the same source, and the cooling frequency ~16 days post-merger is well above the soft X-ray frequencies. Extended Data Figure 4 shows a panchromatic spectrum between the radio and X-ray frequencies. Ultraviolet and near-infrared data are also plotted for comparison. Although the early-time emission in the ultraviolet, optical and infrared frequencies was dominated by thermal emission, at late times there should be a significant synchrotron component. Using the temporal and spectral indices estimated for the radio-only data (earlier in the Methods section), and assuming the cooling break remains beyond $10^{18}$ Hz, we can predict the X-ray flux densities between 0.3-2.2 nJy (flux between $7 \times 10^{-15}$ to $52 \times 10^{-15}$ erg cm$^{-2}$ s$^{-1}$ in the 0.3-10 keV band) on 2017 November 18 (and also for the Chandra observation on December 03-06). *We note that, subsequent to the submission of this paper, the X-ray observations took place and confirmed our prediction*. We estimate the synchrotron cooling frequency as:

For ɣ>>1 (as expected for cocoon):

$$\nu_c \approx 7 \cdot 10^{19} \, \text{Hz} \left(\frac{\gamma}{2}\right)^{-4} \left(\frac{n}{10^{-4} \, \text{cm}^{-3}}\right)^{-3/2} \left(\frac{\epsilon_B}{0.01}\right)^{-3/2} \left(\frac{t}{100 \, \text{days}}\right)^{-2}$$

For β<<1 (i.e. ɣ~1; as expected for the dynamical ejecta tail):

$$\nu_c \approx 2 \cdot 10^{18} \, \text{Hz} \left(\frac{\beta}{0.6}\right)^{-3} \left(\frac{n}{0.03 \, \text{cm}^{-3}}\right)^{-3/2} \left(\frac{\epsilon_B}{0.003}\right)^{-3/2} \left(\frac{t}{100 \, \text{days}}\right)^{-2}$$

We see that the cooling frequency at day ~16 post-merger is much larger than $10^{18}$ Hz, while beyond ~$10^2$-$10^3$ days post-merger this break should be seen moving towards lower frequencies within the electromagnetic spectrum.

**Methods References.**

**Data Availability.** All relevant data are available from the corresponding author on request. Data presented in Figure 1 are included in the Extended Data Table 1.

**Code Availability.** The codes used for generating the synthetic radio light curves are currently being readied for public release (publication in preparation). Radio data processing software: CASA, MIRIAD, DIFMAP.

**Extended Data**

**Extended Data Table 1: Radio data for GW170817**

| UT Date | ΔT (d) | Telescope | ν (GHz) | Bandwidth (GHz) | $S_\nu$ (μJy) |
|---|---|---|---|---|---|
| Sep 16.25 | 29.73 | GMRT | 0.68 | 0.2 | < 246 |
| Sep 17.84 | 31.32 | VLA | 3 | 2 | 34 ± 3.6 |
| Sep 21.86 | 35.34 | VLA | 1.5 | 1 | 44 ± 10 |
| Sep 25.86 | 39.34 | VLA | 15 | 4 | <14.4 |
| Oct 02.79 | 46.26 | VLA | 3 | 2 | 44 ± 4 |
| Oct 09.79 | 53.26 | VLA | 6 | 4 | 32 ± 4 |
| Oct 10.80 | 54.27 | VLA | 3 | 2 | 48 ± 6 |
| Oct 13.75 | 57.22 | VLA | 3 | 2 | 61 ± 9 |
| Oct 21.67 | 65.14 | GMRT | 0.61 | 0.4 | 117 ± 42 |
| Oct 23.69 | 67.16 | VLA | 6 | 4 | 42.6 ± 4.1 |
| Oct 28.73 | 72.20 | VLA | 4.5 | 0.5 | 54.6 ± 5.5 |
| Nov 01.02 | 75.49 | ATCA | 7.25 | 4 | 44.9 ± 5.4 |
| Nov 17.93 | 92.4 | ATCA | 7.25 | 4 | 39.6 ± 7 |
| Nov 18.60 | 93.07 | VLA | 1.6 | 1 | 98 ± 14 |
| Nov 18.66 | 93.13 | VLA | 3 | 2 | 70 ± 5.7 |
| Nov 18.72 | 93.19 | VLA | 15 | 4 | 18.6 ± 3.1 |
| Dec 02.89 | 107.36 | ATCA | 7.25 | 4 | 66.5 ± 5.6 |

Table notes: ΔT represents the time post-merger. The Nov 17.93 ATCA observation was affected by bad weather, and the uncertainty in the flux density is expected to be much larger than the one reported here.

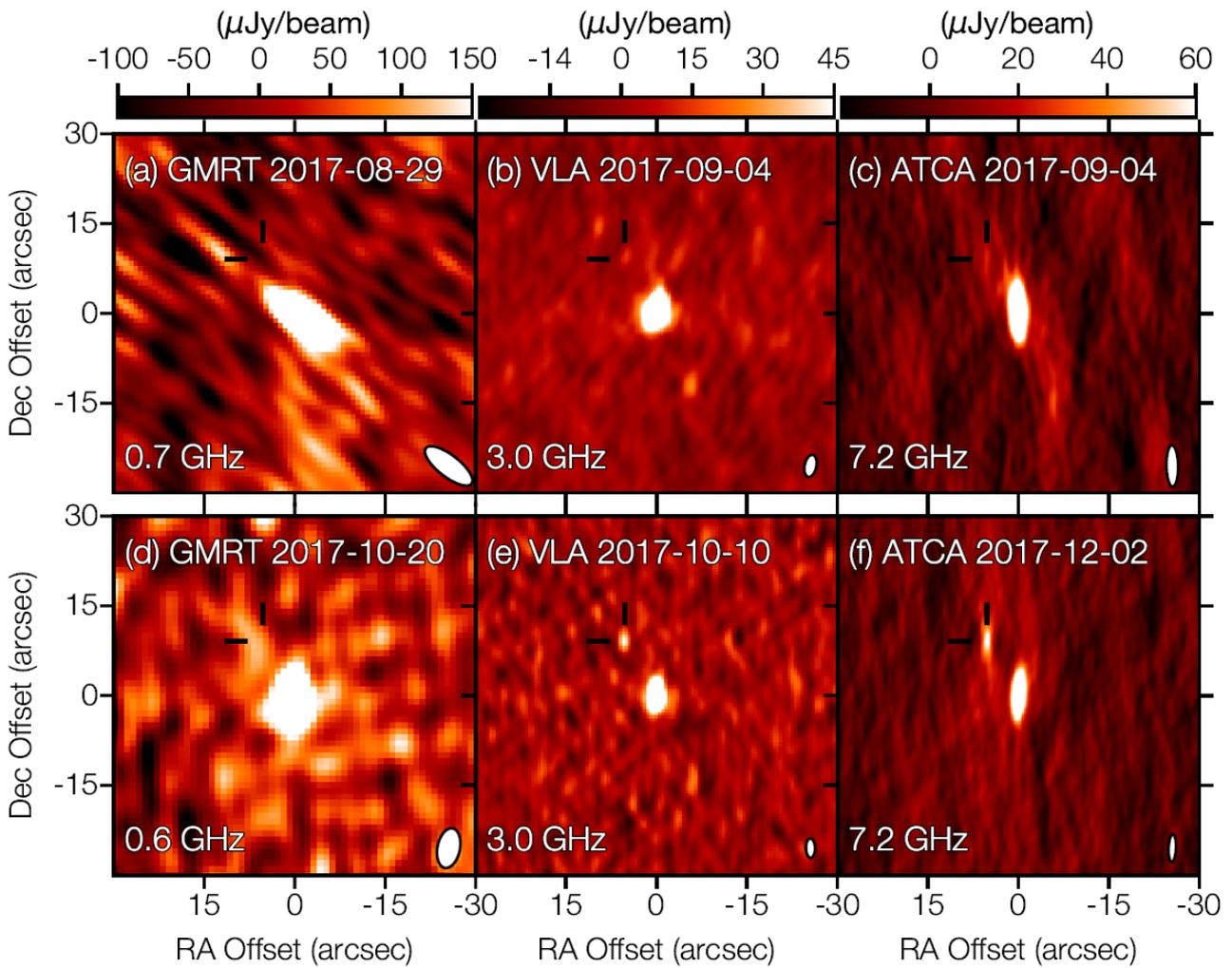

**Extended Data Figure 1. GW170817 radio image cutouts.**

Image cutouts (30" x 30") from the upgraded GMRT, the VLA and the ATCA centred on the NGC 4993. The position of GW170817 is marked by two black lines. Panels (a), (b) and (c) show images from August-September 2017, using the data reported in ref. 12. Panels (d), (e) and (f) show recent data, from October 2017. The flux density scale is denoted by the colorbar in each column. The synthesized beam is shown as an ellipse in the lower right corner of each image.

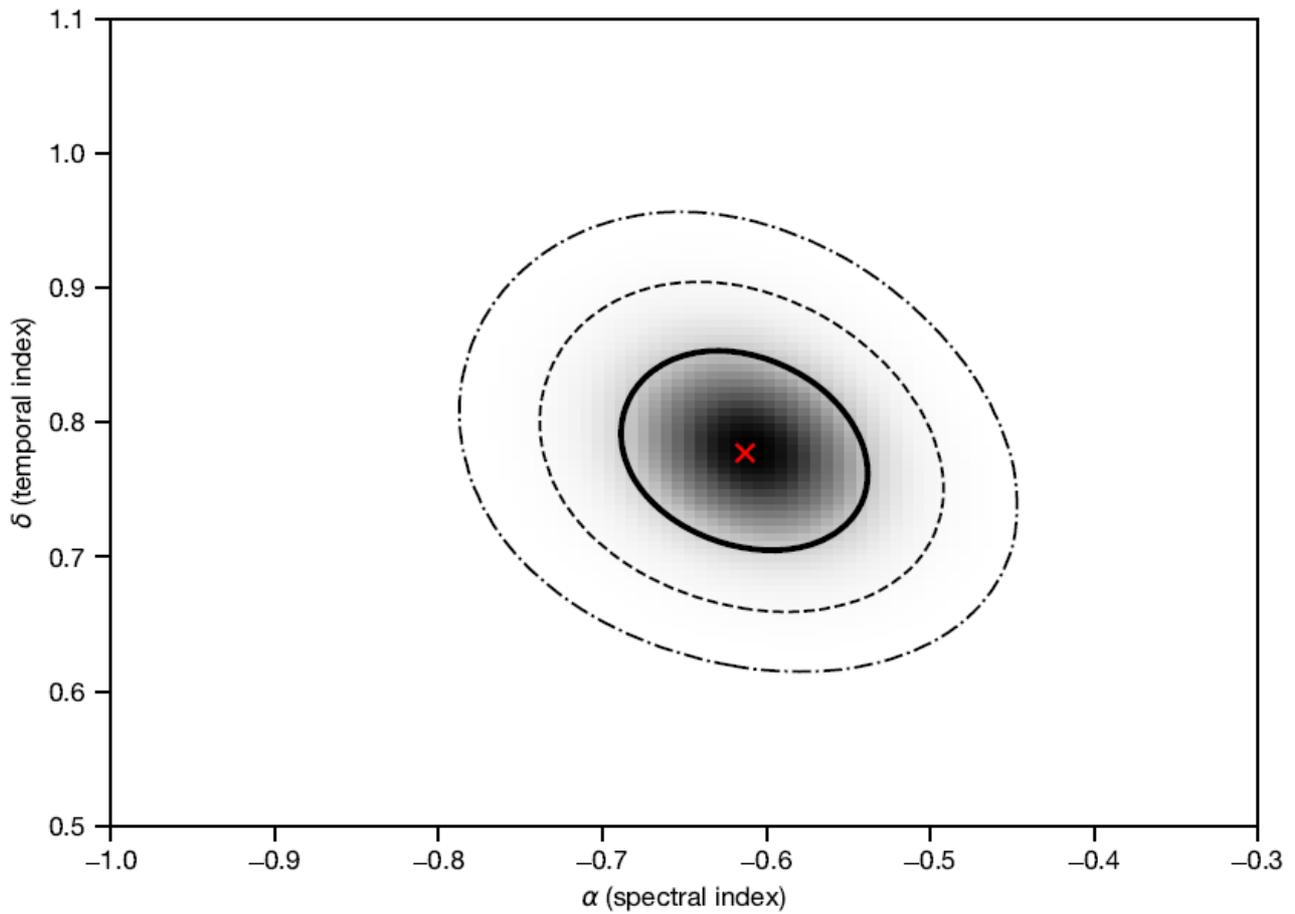

**Extended Data Figure 2. Confidence region for the radio spectral and temporal indices.**

Joint confidence contours for α (the spectral power-law index) and β (the temporal power-law index). The contours are 1-, 2-, and 3-σ confidence contours, and the location of the best-fit values, α=-0.61+/-0.05, δ=0.78+/-0.05, is indicated by the red "x" marker.

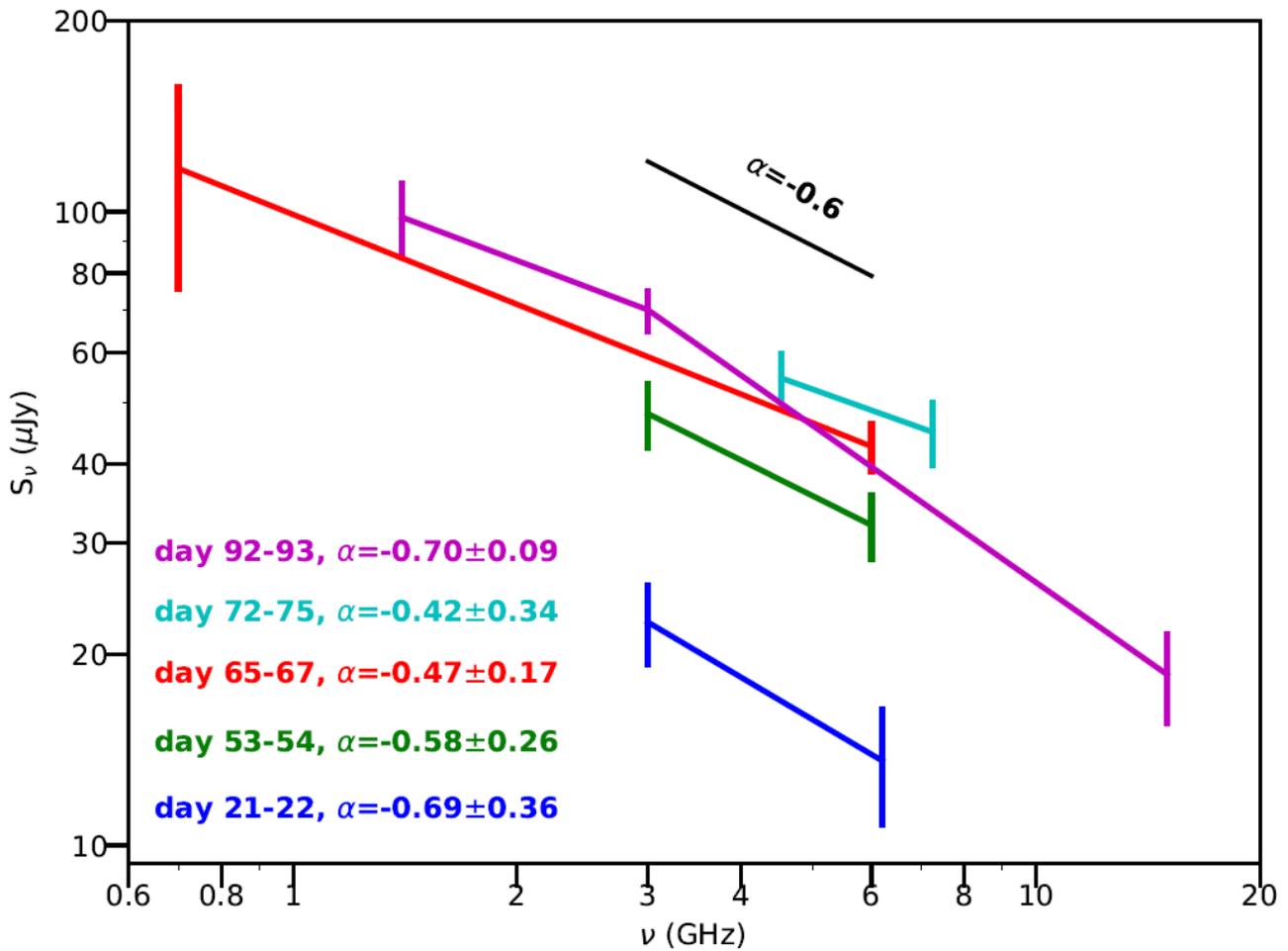

**Extended Data Figure 3**. **Radio-only spectral indices of GW170817.**

Radio spectral indices between 0.6-15 GHz spanning multiple epochs. The different epochs are color coded. The corresponding days post-merger and spectral indices are given in the legend. Error bars are 1σ. The joint analysis of all radio data (see text in Methods section) implies α=-0.61+/-0.05.

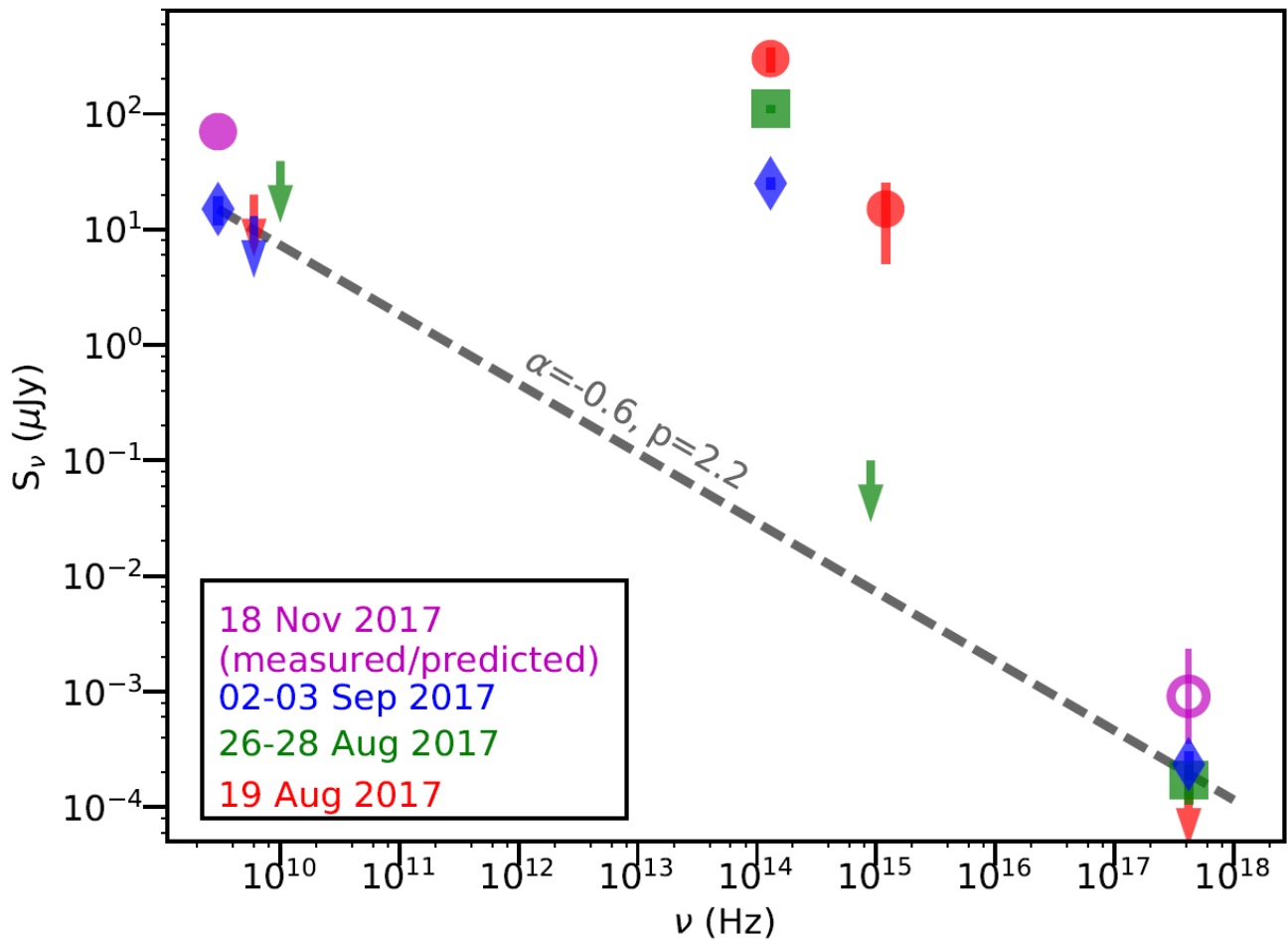

**Extended Data Figure 4. Comparison between the radio and X-ray flux densities of GW170817.**

The comparison of the X-ray data with the radio upper limits (arrows) and detections (filled circles) at different epochs. Error bars are 1σ. The epochs: 2017 August 19, August 26-28, September 02-03 and November 18 (2 days, ~10 days, ~15 days and 93 days post-merger respectively) are color coded (the epoch is given in the legend to the upper-right corner) and marked with different symbols. The spectral index (α) and corresponding electron power law index (p; assuming cooling frequency is beyond $10^{18}$ Hz, as expected for a mildly relativistic outflow) between 3 GHz and $10^{18}$ Hz as derived from the September 02-03 data (α=-0.60+/-0.03 and p=2.20+/-0.06) are consistent with the radio-only spectral indices, and shown here as a dashed grey line. This indicates that the radio and X-rays originate from the same synchrotron source. The corresponding predicted soft-X-ray flux density on November 18 (between 0.3-2.2 nJy; *note: the Chandra X-ray observations from 03-06 December, reported after the submission of this paper, confirmed the prediction*) is shown as a magenta unfilled circle with an error bar. The flux densities in the ultraviolet (~$10^{15}$ Hz) and near-infrared (~$10^{14}$ Hz), dominated by thermal emission at early times, are shown for reference.